# Abrikosov vortex nucleation and its detrimental effect on superconducting spin pumping in Pt/Nb/Ni$_{80}$Fe$_{20}$/Nb/Pt proximity structures

Kun-Rok Jeon,[1,2] Chiara Ciccarelli,[2] Hidekazu Kurebayashi,[3] Lesley F. Cohen,[4] Sachio Komori,[1] Jason W. A. Robinson,[1] and Mark G. Blamire[1]

[1]*Department of Materials Science and Metallurgy, University of Cambridge, 27 Charles Babbage Road, Cambridge CB3 0FS, United Kingdom*

[2]*Cavendish Laboratory, University of Cambridge, Cambridge CB3 0HE, United Kingdom*

[3]*London Centre for Nanotechnology and Department of Electronic and Electrical Engineering at University of College London, London WC1H 01H, United Kingdom*

[4]*The Blackett Laboratory, Imperial College London, SW7 2AZ, United Kingdom*

**We report Abrikosov vortex nucleation in Pt/Nb/Ni$_{80}$Fe$_{20}$/Nb/Pt proximity-coupled structures under oblique ferromagnetic resonance (FMR) that turns out to be detrimental to superconducting spin pumping. By measuring an out-of-plane field-angle $\Theta_H$ dependence and comparison with Pt-absent control samples, we show that as $\Theta_H$ increases, the degree of enhancement (suppression) of spin pumping efficiency in the superconducting state for the Pt-present (Pt-absent) sample diminishes and it reverts to the normal state value at $\Theta_H = 90°$. This can be explained in terms of a substantial out-of-plane component of the resonance field for the Ni$_{80}$Fe$_{20}$ layer (with in-plane magnetization anisotropy and high aspect ratio) that approaches the upper critical field of the Nb, turning a large fraction of the singlet superconductor volume into the normal state.**

As described by the Ginzburg-Landau (GL) theory in 1950 [1], the response of superconducting materials to an applied magnetic field depends on the value of GL parameter $\kappa$, denoted by the ratio of the London penetration depth $\lambda_L$ to the superconducting coherence length $\xi_{SC}$. This material parameter classifies superconductors (SCs) into two categories: type-I SCs when $\kappa < 1/\sqrt{2} = 0.71$ and type-II SCs when $\kappa > 0.71$ [2,3].

A type-I SC ($\kappa < 0.71$) under a magnetic field smaller than the thermodynamic critical field $\mu_0 H_c$ expels the magnetic field from its interior, except in thin boundary layers (known as the Meissner state). For an applied field larger than $\mu_0 H_c$, superconductivity is abruptly destroyed, and the SC is in the normal state, fully penetrated by the magnetic field. By contrast, a type-II SC ($\kappa > 0.71$) energetically favors to split into as many domains as possible because of the negative wall energy of a non-superconducting (normal state) domain and a superconducting (Meissner state) domain [1-3]. This results in the existence of mixed state or unstable superconducting state for an intermediate magnetic field between the lower and upper critical fields ($\mu_0 H_{c1}$ and $\mu_0 H_{c2}$, respectively), where the magnetic field can partially penetrate the type-II SC in the form of Abrikosov vortices (also called flux lines, flux tubes, or fluxons) each carrying a quantum of magnetic flux $\Phi_0 = h/2e = 2.07 \times 10^{-15}$ T·m$^2$ [4]. As the magnetic field increases from $\mu_0 H_{c1}$ to $\mu_0 H_{c2}$, more and more flux lines penetrate and the density of Abrikosov vortices grows. When reaching $\mu_0 H_{c2}$, the normal state vortex cores overlap completely such that the superconducting volume fraction shrinks down to zero and the superconductivity vanishes.

Early studies based on electromagnetic interaction or magnetic stray fields (i.e. the orbital effect) in type-II SC and ferromagnet (FM) hybrid structures [5,6] have found several interesting phenomena involving Abrikosov vortices: (reverse) domain-wall superconductivity [7,8], vortex pinning by magnetic objects [9,10], and spontaneous formation of vortex-antivortex pairs [11-14]. A very recent experiment of demonstrating magnon-fluxon interaction in a FM/SC heterostructure [15] has drawn renewed interest in this research direction. Note that in all these schemes, SC and FM are spatially separated by a thin insulating layer and thus no proximity effect is present.

In recent years, it has become clear that SC/FM *proximity-coupled* structures can create new physical phenomena whose unique properties can greatly improve central effects in the field of spintronics [16-21]. For example, equal-spin triplet Cooper pairs, generated via spin mixing and spin rotation processes at engineered magnetically-inhomogeneous SC/FM interfaces [19-21], can carry non-dissipative spin angular momentum in equilibrium (ground-state) nature. Although there have been quite recent works [22,23] which focus on vortex liquid phase (or vortex flow) with a non-zero resistance activated for rather high magnetic fields near $\mu_0 H_{c2}$ in insulating FM/thick SC systems, our understanding of the generic role of Abrikosov vortices in proximity-coupled systems remains in the initial stage; especially for magnetization dynamics and spin transport [24-26].

Here, we present out-of-plane (OOP) field-angle $\theta_H$ dependence of ferromagnetic resonance (FMR) measurements on two types of Nb(30 nm)/Ni$_{80}$Fe$_{20}$(6 nm)/Nb(30 nm) proximity structures with and without Pt(5 nm) layers [Figs. 1(a) and 1(b)] to *intentionally* nucleate OOP vortices in the Nb (type-II SC) layers [2,3] and to investigate their influence on spin pumping phenomena in the superconducting state [26]. We note that in such structures where the Nb thickness $t_{Nb}$ is far below $\lambda_L$ of Nb thin films ($\geq$ 100 nm) [27], the emergence of in-plane (IP) vortices is energetically unfavorable because their Gibbs free energy is higher than that of OOP vortices by a factor of $(\lambda_L/t_{Nb})^2$ [3,28,29]. Combined measurements of $\theta_H$-dependent FMR spectra and static magnetic properties of the samples (with and without the Pt layers) show that a larger vortex population at a higher $\theta_H$ FMR remarkably reduces the degree of change in spin pumping efficiency across the superconducting transition temperature $T_c$. This can be straightforwardly interpreted due to a larger number of OOP vortices (nucleated at a higher $\theta_H$ with a stronger resonance field) that reduce profoundly the real superconducting volume, the effective pair potential, and so the overall singlet pair density which is the underlying source of proximity-induced triplet pairing.

For experimental details, sample preparation and measurement setup have been described previously [26,29]. Here we only describe the essential role of Pt layers in our Pt/Nb/Ni$_{80}$Fe$_{20}$/Nb/Pt proximity system [26,30]. When the Pt layers are absent, the diffusion of pumped non-equilibrium spin accumulation from the precessing Ni$_{80}$Fe$_{20}$ into the Nb layers is precluded by the opening of the spin-singlet superconducting gap in the density of sates (DOS) below $T_c$ [24-26]. However, in the presence of the Pt, equal-spin triplet states are proximity-induced into the Nb (singlet SC) layers by spin-orbit coupling (SOC) in concert with exchange field [31,32], which turns the Nb/Pt composite layers to an efficient superconducting spin sink and consequently leads to the greatly enhanced spin pumping/transfer beyond the normal state - this can be probed by FMR linewidth broadening or Gilbert damping increase [26,33].

Figure 2 shows typical FMR spectra [26,29] of Nb/Ni$_{80}$Fe$_{20}$/Nb samples with and without Pt layers taken at selected values of $\theta_H$ at a fixed microwave frequency $f$ = 10 GHz, above and below $T_c$ of the coupled Nb layers. We determine the FMR linewidth $\mu_0\Delta H$ [linked to the (effective) Gilbert damping $\alpha$] and the resonance magnetic field $\mu_0 H_{res}$ [associated with the (effective) saturation magnetization $\mu_0 M_{eff}$] by fitting the field derivative of symmetric and antisymmetric Lorentzian functions (black solid lines in Fig. 1) to the FMR data [34].

The extracted values of $\mu_0 H_{res}$ and $\mu_0\Delta H$ are plotted as a function of $\theta_H$ in Fig. 3 from which one can obtain the (effective) Gilbert damping $\alpha$ for a given fixed $f$, according to the model developed earlier [35,36]:

$$\sin(2\theta_M) = (2\mu_0 H_{res}/\mu_0 M_{eff}) \cdot \sin(\theta_H - \theta_M), \quad (1)$$

$$f = \frac{\gamma}{2\pi}\sqrt{\mu_0 H_1(\theta_H, \theta_M) \cdot \mu_0 H_2(\theta_H, \theta_M)}, \quad (2)$$

$$\mu_0 H_1(\theta_H, \theta_M) = \mu_0 H_{res} \cdot \cos(\theta_H - \theta_M) - \mu_0 M_{eff} \cdot \sin^2(\theta_M), \quad (3)$$

$$\mu_0 H_2(\theta_H, \theta_M) = \mu_0 H_{res} \cdot \cos(\theta_H - \theta_M) + \mu_0 M_{eff} \cdot \cos(2\theta_M). \quad (4)$$

Here $\theta_M$ is the OOP magnetization-angle of the $Ni_{80}Fe_{20}$ layer and we take the case for which the IP magnetization-angle $\phi_M$ of FM and the IP field-angle $\phi_H$ are collinearly aligned [$\phi_M = \phi_H$, see Fig. 1(b)], as relevant to our experimental setup. Using Eq. (1) with the extracted $\mu_0 M_{eff}$ value from the $f$-dependent IP FMR spectra ($\theta_H = 0°$, see the Supplementary Material [37] for details) and the measured $\mu_0 H_{res}(\theta_H)$ value, we obtain the $\theta_H$ dependence of $\theta_M$ [insets of Figs. 3(a) and 3(c)].

The measured total linewidth [Figs. 3(b) and 3(d)] can be decomposed into (1) the intrinsic contribution $\mu_0\Delta H_{int}$ due to the (viscous) damping of precessing magnetization $\alpha$ [36,38] and (2) the extrinsic contribution $\mu_0\Delta H_{ext}$ resulting from the magnetic inhomogeneities $\mu_0\Delta H_{inhom}$ [39] and the two-magnon scattering (TMS) $\mu_0\Delta H_{TMS}$ [40-42]:

$$\mu_0\Delta H = \mu_0\Delta H_{int} + \mu_0\Delta H_{ext} \approx \mu_0\Delta H_{int} + (\mu_0\Delta H_{inhom} + \mu_0\Delta H_{TMS}), \quad (5)$$

$$\mu_0 \Delta H_{int} = \frac{1}{\sqrt{3}} \alpha \cdot [\mu_0 H_1(\theta_H, \theta_M) + \mu_0 H_2(\theta_H, \theta_M)] \cdot \left| \frac{d[\frac{2\pi f(\theta_H, \theta_M)}{\gamma}]}{d[\mu_0 H_{res}]} \right|^{-1}. \quad (6)$$

Assuming $\mu_0 \Delta H \approx \mu_0 \Delta H_{int}$, a first-order estimate of $\alpha$ can be available from the $\mu_0 \Delta H(\theta_H)$ data by using Eq. (6) with the values of $\theta_M(\theta_H)$ and $\mu_0 H_{res}(\theta_H)$ [solid lines in Figs. 3(b) and 3(d)]. Here we treat $\alpha$ as a single adjustable parameter. The estimated $\alpha$ values are consistent with those obtained from $f$-dependent FMR spectra at $\theta_H = 0°$ (see Ref. [37]), implying that $\mu_0 \Delta H_{ext}$ has a minor contribution to the total linewidth of our samples at $f$ = 10 GHz. We then find the suppressed and enhanced FMR damping in the superconducting state for the Pt-absent and Pt-present samples, respectively, compared each with its normal state value [Figs. 3(b) and 3(d)]. This superconducting state enhancement of FMR damping relevant to the presence of Pt layers can be interpreted in terms of the proximity generation of spin-polarized triplet pairs via SOC at the Nb/Pt interface, acting in conjunction with a non-locally supplied exchange field [25,43], as described above. Our recent experiment [44], proving the explicit correlation of superconducting spin pumping efficiency with the strength of Fe-induced direct exchange field at the Nb/Pt/(Fe) interface, also supports this interpretation.

Let us now focus on the key aspect of the $\mu_0 \Delta H(\theta_H)$ data in Figs. 3(b) and 3(d). When $\theta_H$ is larger than 60°, requiring a substantial $\mu_0 H_{res}$ (> 0.3 T) to rotate the magnetization precession axis of IP magnetized $Ni_{80}Fe_{20}$ to the field direction [Figs. 3(a) and 3(c)], the superconducting state broadening (narrowing) of $\mu_0 \Delta H$ for the Pt-present (Pt-absent) sample diminishes considerably and it returns to the normal state value at $\theta_H \approx 90°$.

We first discuss the contribution of $\theta_H$-dependent $\mu_0 \Delta H_{ext}$ to the total linewidth. The $f$-dependent FMR data obtained at $\theta_H = 0°$ [e.g. small $\mu_0 \Delta H_{ext}$ ($\leq |0.5$ mT|) and linear

$f$-dependence, see Ref. [37]] indicates that our samples are of high quality and basically free from TMS. In addition, for a thin FM with defects, this TMS process [i.e. the defect-mediated coupling of the uniform precessional mode ($k_\parallel = 0$) to degenerate finite-$k$ ($k_\parallel \neq 0$) spin-wave modes] is known to be activated for $\theta_M < 45°$ when finite-$k$ modes equi-energy with the uniform FMR mode are present [40-42]. To activate the TMS, a term in the spin-wave dispersion relation $f^2(k_\parallel)$ linear in $k_\parallel$ has negative coefficient, or equivalently the initial slope of $f^2(k_\parallel)$ is negative [45,46]:

$$f^2(k_\parallel) \approx f^2 - \frac{\gamma^2}{8\pi^2} \mu_0 M_{eff} k_\parallel t_{FM} \cdot \{\mu_0 H_1(\theta_H, \theta_M) \cdot [cos^2(\theta_M) - sin^2(\theta_M) \cdot cos^2(\phi_{k_\parallel})] - H_2(\theta_H, \theta_M) \cdot sin^2(\phi_{k_\parallel})\} + \frac{\gamma^2}{4\pi^2} D k_\parallel^2 \{[\mu_0 H_1(\theta_H, \theta_M) + \mu_0 H_2(\theta_H, \theta_M)\}$$

(7)

Here $\phi_{k_\parallel}$ is the direction of propagation of the spin-wave in the film plane relative to the IP projection of the magnetization ($\phi_M$), $t_{FM}$ is the $Ni_{80}Fe_{20}$ thickness (6 nm), and $D$ is the $Ni_{80}Fe_{20}$ exchange stiffness ($\sim 10^{-17}$ T·m$^2$). Since the calculation results using Eq. (7) [insets of Figs. 3(b) and 3(d)] predict that the initial slope of $f^2(k_\parallel)$ becomes positive when $\theta_H \geq 80°$ ($\theta_M > 45°$) for our samples; thereby vanishing of spin-wave modes degenerate with the FMR mode at $\theta_H = 80° - 90°$, we rule out the TMS mechanism.

Next, we show that OOP vortex nucleation in the Nb layers, reducing the active volume of (singlet) superconducting domains as well as the effective pair potential, is responsible for the observed high $\theta_H$ behavior. This can more readily be seen by plotting the superconducting gap $2\Delta$, $\mu_0 H_{res}$, and $\mu_0 \Delta H$ versus the normalized temperature $T/T_c$ for four different $\theta_H$ (Fig. 4). In these plots, the $T/T_c$-dependent $2\Delta(\theta_H)$ is calculated from the measured $T_c(\theta_H)$ under FMR condition [inset of Fig. 4(a) and 4(d)] [3], and the measured $\mu_0 H_{res}$ and $\mu_0 \Delta H$ values are normalized each to its normal state one at 8 K for quantitative analysis.

Upon entering the superconducting state ($T/T_c < 1$), $\mu_0 H_{res}(T/T_c)$ remains almost insensitive to $\theta_H$ [Figs. 4(b) and 4(e)] whereas a significant $\theta_H$-dependent evolution of $\mu_0 \Delta H(T/T_c)$ appears [Figs. 4(c) and 4(f)]: a visible diminishment of the broadening (narrowing) of $\mu_0 \Delta H$ for the Pt-present (Pt-absent) sample with the increase of $\theta_H$ from $0^0$ to $90^0$. Most importantly, we can see in the $2\Delta(T/T_c)$ and $\mu_0 \Delta H(T/T_c)$ plots that the absolute magnitude of change in spin pumping efficiency across $T_c$ is positively correlated with the effective pair potential of the Nb layers, linked to the real superconducting volume, for both types of the samples. In fact, this result agrees well with general understanding of spin-triplet proximity effect in that (equal-spin) triplet proximity pairing necessary for spin angular momentum transfer relies on the strength of the underlying singlet superconductivity (i.e. the singlet pair density) [16-21] and with the previous experiments on OOP triplet spin valves [47,48].

One can, in principle, calculate the active volume $V_{SC}^{cal}$ of superconducting domains in Nb films under the OOP applied field $\mu_0 H_\perp$ [2,3]:

$$V_{SC}^{cal} \approx \left[1 - \frac{\pi \cdot (\xi_{SC})^2}{(a_{VL})^2}\right], \quad (8)$$

where $\xi_{SC} = \xi(0)/[1 - T/T_c]^{1/2}$, $\xi(0)$ is the zero-$T$ coherence length of the Nb films (~13 nm) in the dirty limit, $a_{VL} = \left[2\Phi_0/\sqrt{3}\mu_0 H_\perp\right]^{1/2}$ is the vortex lattice parameter, and $\mu_0 H_\perp = \mu_0 H_{res} \cdot \sin(\theta_H) > \mu_0 H_{c1\perp}$. Note that $V_{SC}^{cal} < 0$ means the collapse of superconductivity due to overlapping non-superconducting (normal-state) vortex cores. As summarized in Table I, $V_{SC}^{cal}$ at 2 K is predicted to be much smaller as $\theta_H$ approaches 90° for both types of the samples - this, along with $2\Delta(\theta_H)$ at 2 K (Table I), basically explains the experimental observation and captures the underlying mechanism, that is, the OOP vortex nucleation deteriorating the (singlet) superconductivity.

Finally, we measure static magnetic properties of the samples with and without the Pt layers across $T_c$ by applying $\mu_0 H$ at $\theta_H = 0°$, 45°, 70°, and 90° (Fig. 5). From the Nb magnetization curve $M_{Nb}(\mu_0 H)$ (insets of Fig. 5), isolated by taking the difference between the total magnetization curves (of the sample) above and below $T_c$ [29], we can ensure that for $\theta_H = 0°$ ($\theta_H = 90°$), FMR is taken far below (in the vicinity of) $\mu_0 H_{c2}$ of $M_{Nb}(\mu_0 H)$ over which the singlet superconductivity is completed destroyed and so the Nb layers are in the normal state. Furthermore, the superconducting volume fractions $V_{SC}^{mea}(\theta_H) \approx \left[1 - \frac{\mu_0 H_{res}(\theta_H)}{\mu_0 H_{c2}(\theta_H)}\right]$ extracted from the measured $M_{Nb}(\mu_0 H)$ curves at four different $\theta_H$ (insets of Fig. 5) are in reasonable agreement with the $V_{SC}^{cal}$ values using Eq. (8) (Table I), strongly supporting our claim.

In conclusion, we investigate how Arbrikosov vortex nucleation influences superconducting spin pumping in Pt/Nb/Ni$_{80}$Fe$_{20}$/Nb/Pt proximity structures by measuring the $\theta_H$ dependence of FMR spectra and comparison with Pt-absent (control) samples. We clarify that the degree of change in spin pumping efficiency across $T_c$ is positively correlated with the effective pair potential, the real superconducting volume, and thus the singlet pair density of the Nb layers which is the underlying source of proximity-induced triplet pairing. As a result, a larger OOP vortex population nucleated at a higher $\theta_H$ FMR (with a stronger $\mu_0 H_{res}$) turns out to be detrimental to the generation of superconducting spin currents. Our work highlights the importance of circumventing the unintentional (OOP) vortex nucleation for more efficient conversion of spin-singlets to equal-spin triplets in SC/FM proximity systems [16-21].

We would like to thank Xavier Montiel and Matthias Eschrig for valuable

discussions. This work was supported by EPSRC Programme Grant No. EP/N017242/1.

**Figure captions**

FIG. 1. (a) Schematic of Pt($t_{Pt}$)/Nb(30 nm)/Ni$_{80}$Fe$_{20}$(6 nm)/Nb(30 nm)/Pt($t_{Pt}$) proximity structures with two different Pt thicknesses $t_{Pt}$ of 0 and 5 nm. (b) Coordinate system used in the present FMR study.

FIG. 2. (a) Representative FMR spectra of the Nb(30 nm)/Ni$_{80}$Fe$_{20}$(6 nm)/Nb(30 nm) control sample taken at various values of OOP field-angle $\Theta_H$ at a fixed microwave frequency $f$ = 10 GHz, above (yellow background) and below (blue background) the superconducting transition $T_c$ of the Nb. (b) Data equivalent to (a) but for the Pt(5 nm)/Nb(30 nm)/Ni$_{80}$Fe$_{20}$(6 nm)/Nb(30 nm)/Pt(5 nm) sample. The black solid lines are fits to precisely determine the FMR linewidth $\mu_0\Delta H$ and the resonance magnetic field $\mu_0 H_{res}$ [34].

FIG. 3. (a) Resonance magnetic field $\mu_0 H_{res}$ and (b) FMR linewidth $\mu_0\Delta H$ as a function of OOP field-angle $\Theta_H$ for the Nb(30 nm)/Ni$_{80}$Fe$_{20}$(6 nm)/Nb(30 nm) control sample. The upper inset shows the calculated OOP magnetization-angle $\Theta_M$ versus the applied OOP field-angle $\Theta_H$; the lower inset displays the derived spin-wave dispersion $f^2(k_\parallel)$ from Eq. (7) at $f$ = 10 GHz at $\phi_{k_\parallel}$ = 0º. (c),(d) Data equivalent to (a),(b) but for the Pt(5 nm)/Nb(30 nm)/Ni$_{80}$Fe$_{20}$(6 nm)/Nb(30 nm)/Pt(5 nm) sample. The solid lines are fits to deduce the (effective) Gilbert damping constant $\alpha$ using Eq. (6). Note that a slight asymmetry in the $\Theta_H$-dependent FMR data with respect to $\Theta_H = 90°$ is caused by unintentional misalignment between the film plane and the applied field (less than 3°).

FIG. 4. (a) Superconducting gap $2\Delta$, (b) FMR linewidth $\mu_0 \Delta H$, and (c) resonance field $\mu_0 H_{res}$ as a function of normalized temperature $T/T_c$ for the Nb(30 nm)/Ni$_{80}$Fe$_{20}$(6 nm)/Nb(30 nm) control sample, obtained at four different values of OOP field-angle $\Theta_H$. Note that the $2\Delta$ is calculated from the measured $T_c$ under FMR condition [inset: $dR(T)/dT$] [3] and the $\mu_0 \Delta H$ and $\mu_0 H_{res}$ values are normalized each to its normal state one at 8 K. (d)-(f) Data equivalent to (a)-(c) but for the Pt(5 nm)/Nb(30 nm)/Ni$_{80}$Fe$_{20}$(6 nm)/Nb(30 nm)/Pt(5 nm) sample.

FIG. 5. (a) Total magnetization $M_{tot}$ versus magnetic field $\mu_0 H$ curves for the Nb(30 nm)/Ni$_{80}$Fe$_{20}$(6 nm)/Nb(30 nm) control sample, obtained by applying $\mu_0 H$ at $\theta_H = 0°, 45°, 70°$, and $90°$ at the temperature $T$ of 2 and 8 K. The diamagnetic background signal from the sample holder is subtracted. Each inset shows the isolated Nb (type-II SC) magnetization $M_{Nb}(\mu_0 H)$ by taking the difference between $M_{tot}(\mu_0 H)$ curves above and below the superconducting transition $T_c$ of the Nb layers [29]. (b) Data equivalent to (a) but for the Pt(5 nm)/Nb(30 nm)/Ni$_{80}$Fe$_{20}$(6 nm)/Nb(30 nm)/Pt(5 nm) sample. Note that the ratio of IP and OOP components of $M_{Nb}(\theta_H)$ is given approximately by $\tan(\theta_H)/(1-D)$, where $D$ is the demagnetization factor [3]. Thus at $\theta_H \neq 0$, $M_{Nb}(\theta_H)$ is dominated by the OOP component due to the large value of $D \approx 1$ for our sample geometry [3].

TABLE I. Calculated (measured) active volume $V_{SC}^{cal}$ ($V_{SC}^{mea}$) of superconducting

domains in the two types of the samples from Eq. (8) (from Fig. 5) at 2 K for four different values of OOP field-angle $\varTheta_\mathrm{H}$, along with the corresponding (effective) pair potential $2\Delta$ [from Figs. 4(a) and 4(c)]. Note that a large error in $V_{SC}^{mea}$ is due to the uncertainty in determination of $H_{c2}$ from $M_\mathrm{Nb}(\mu_0 H)$ curves (insets of Fig. 5).

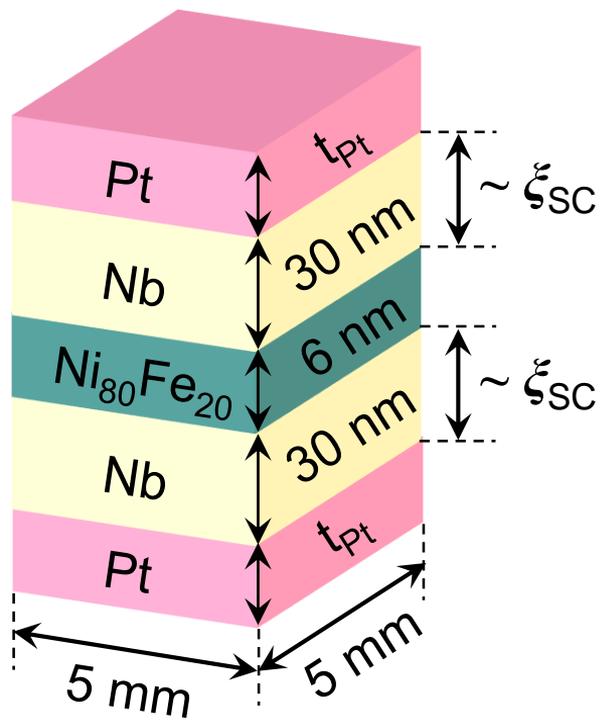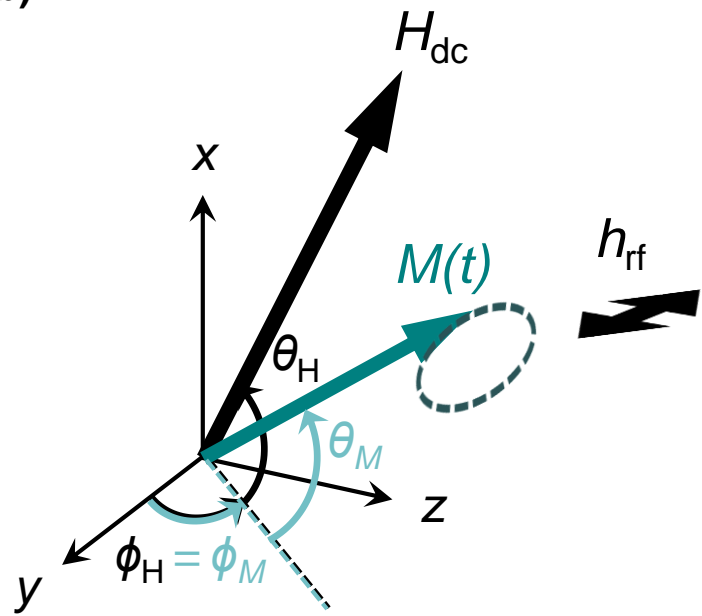

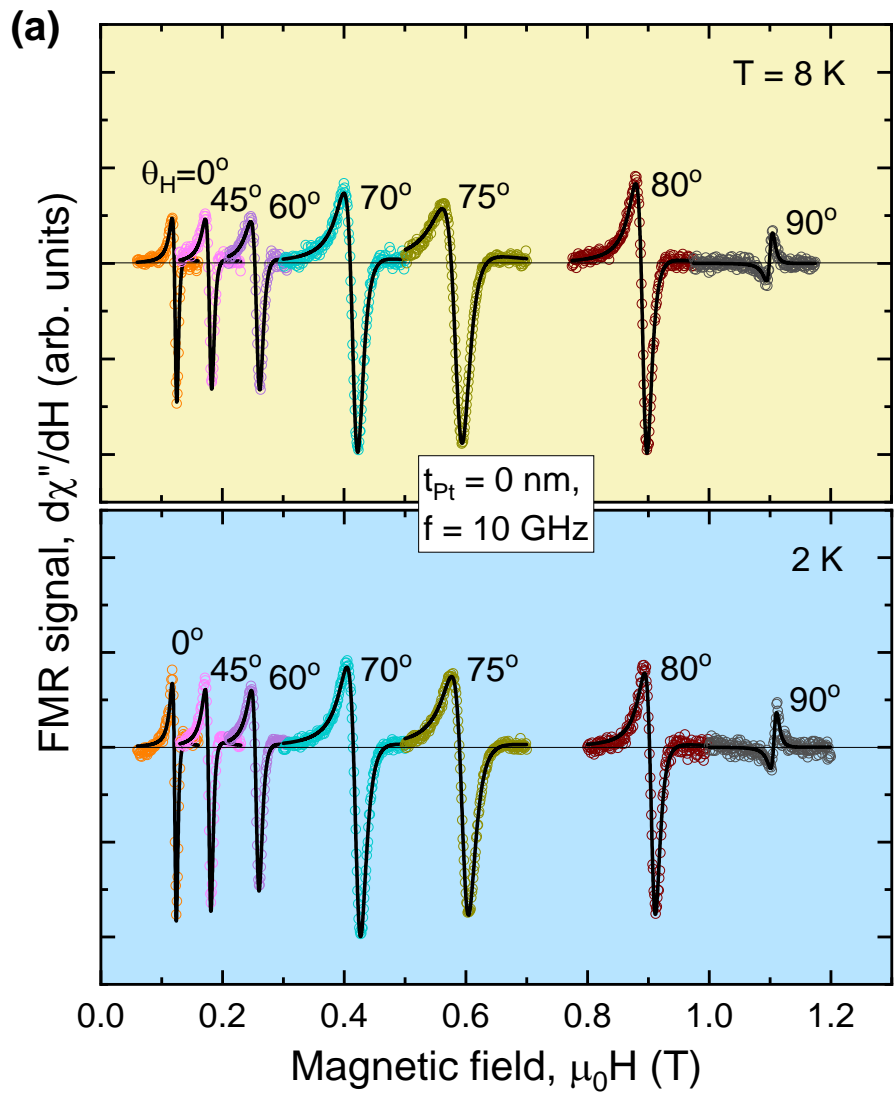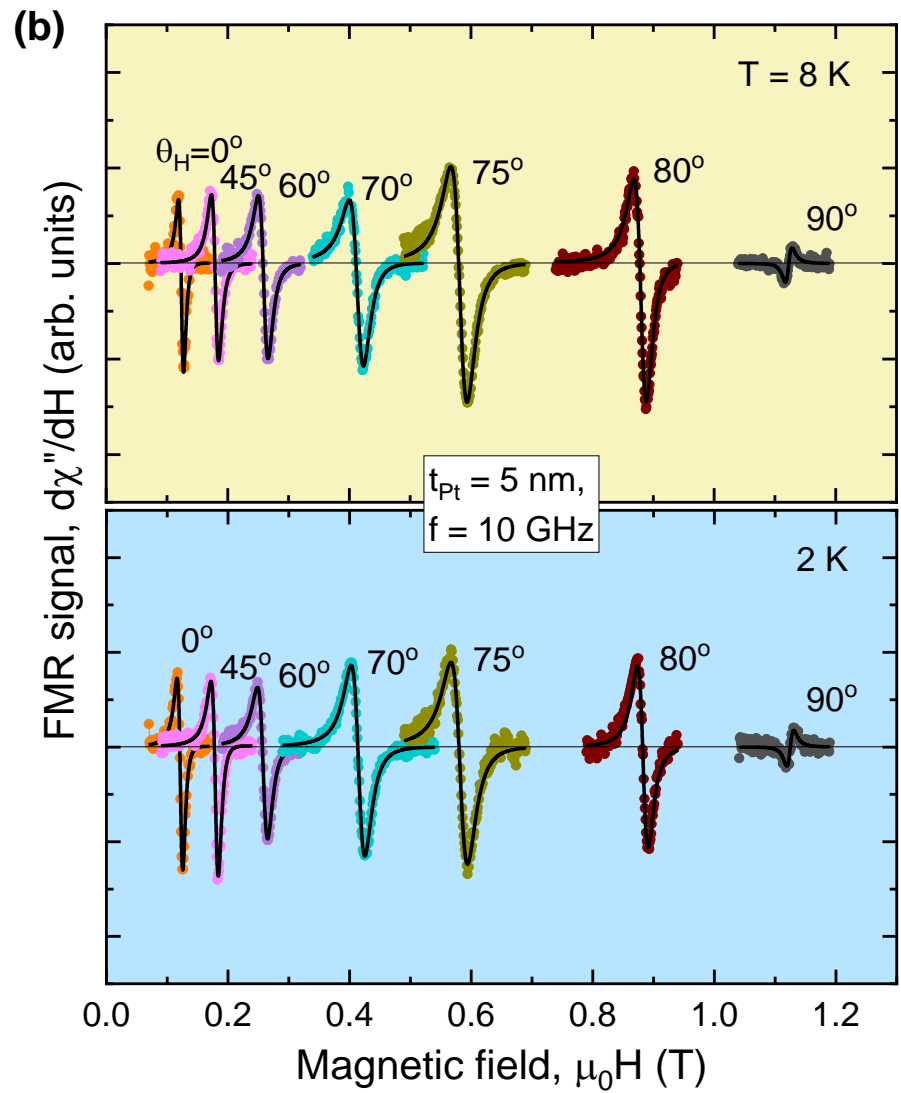

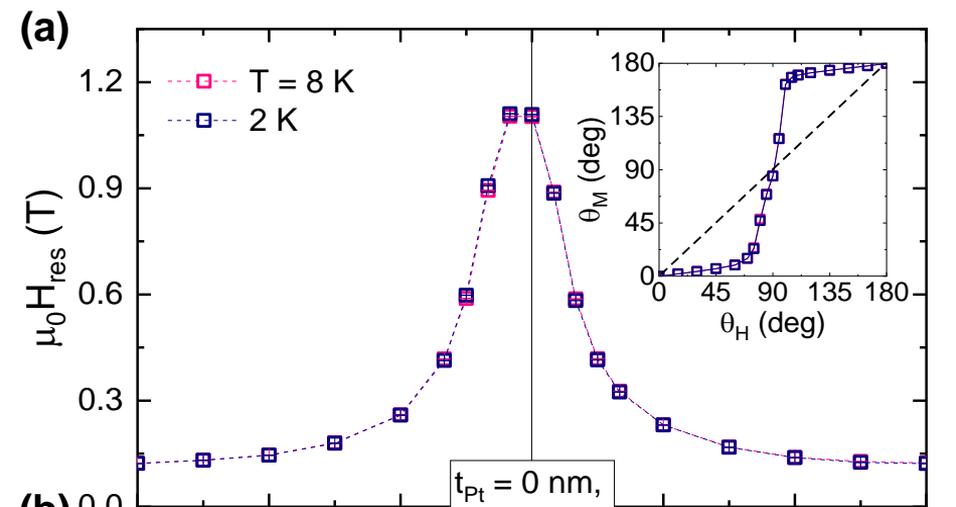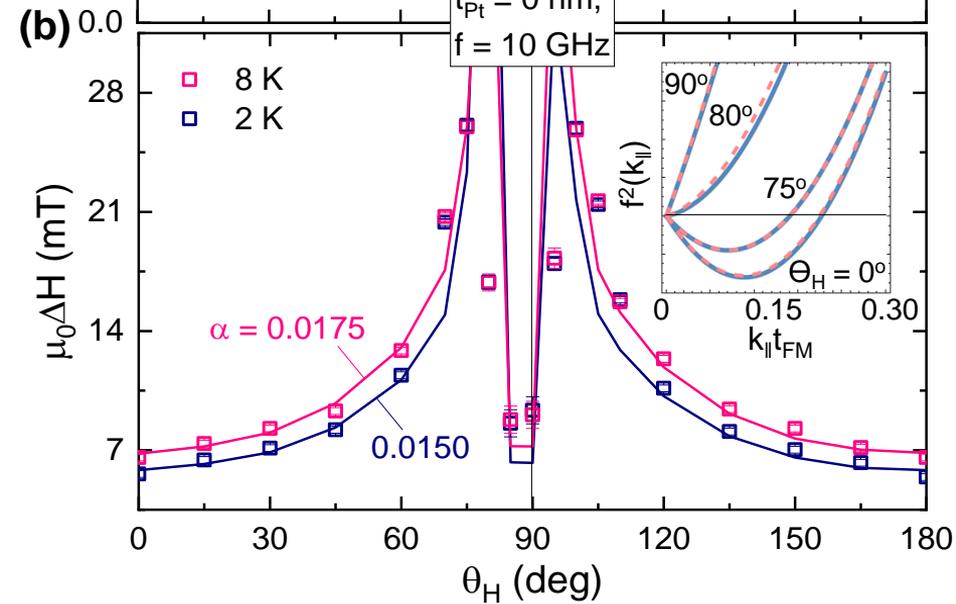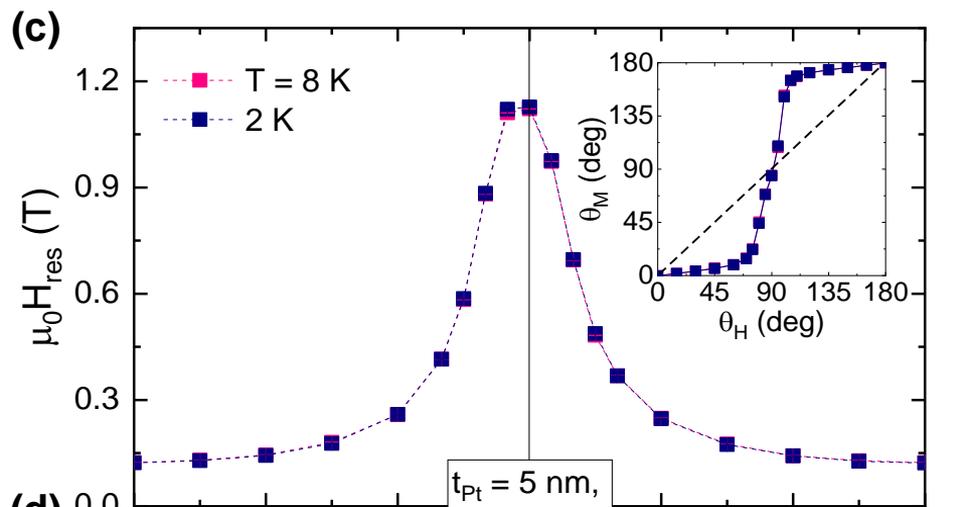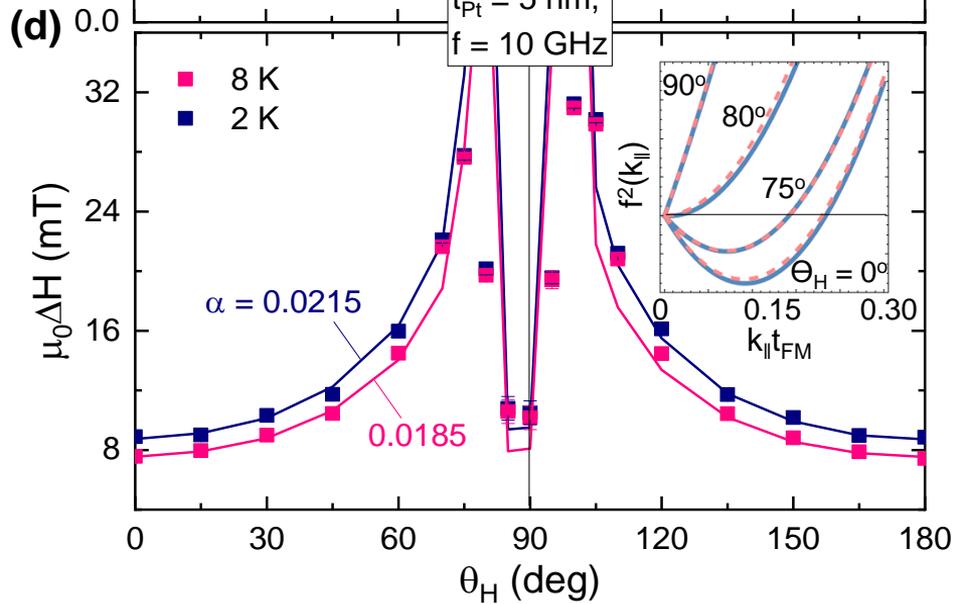

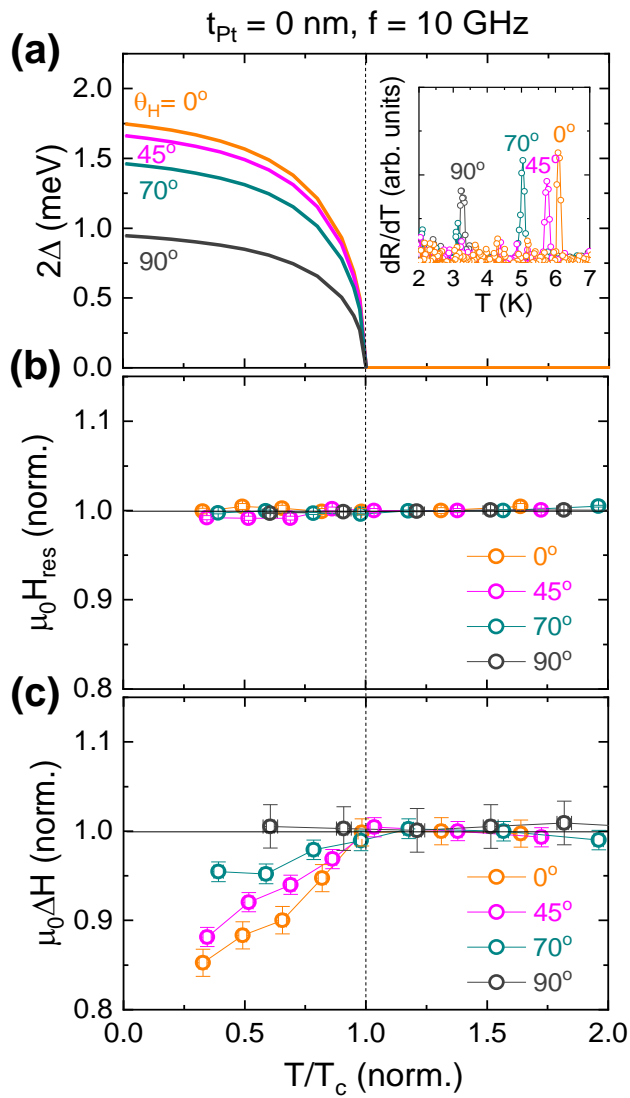
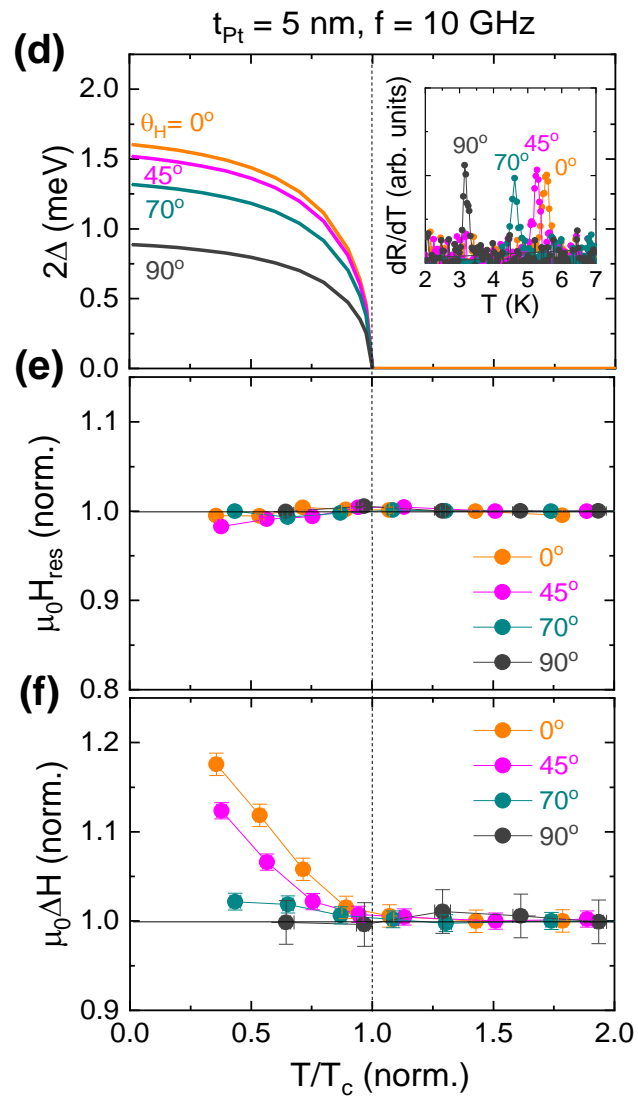

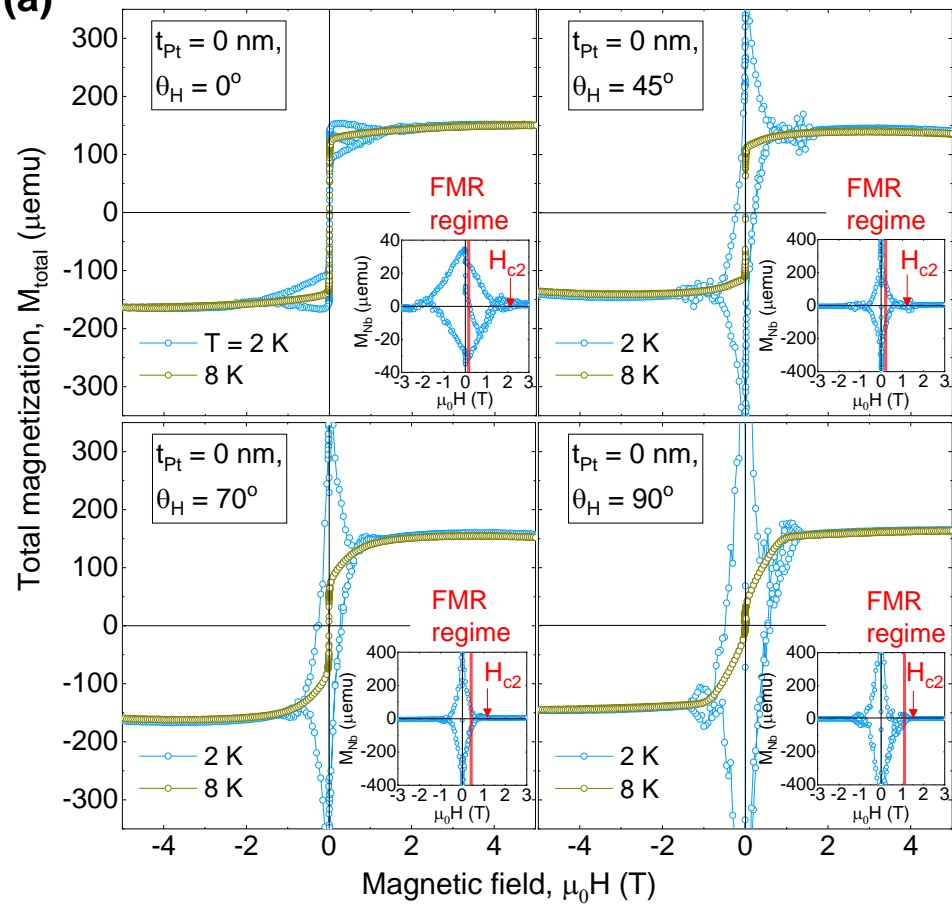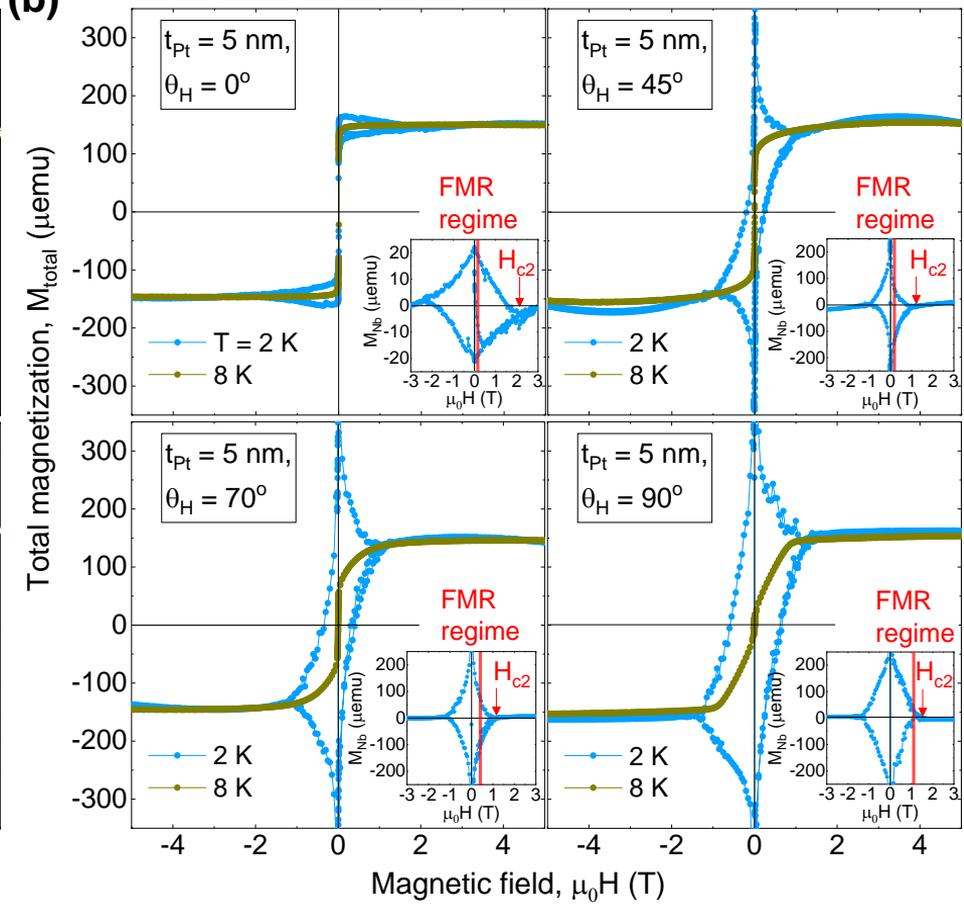

Table I

| $\theta_H$ | No Pt | | | With Pt | | |
|---|---|---|---|---|---|---|
| | $V_{SC}^{cal}$ [%] | $V_{SC}^{mea}$ [%] | $2\Delta$ [meV] | $V_{SC}^{cal}$ [%] | $V_{SC}^{mea}$ [%] | $2\Delta$ [meV] |
| $0^0$ | 100 | 95 ± 2 | 1.65 | 100 | 94 ± 2 | 1.51 |
| $45^0$ | 96 | 91 ± 3 | 1.57 | 95 | 90 ± 3 | 1.42 |
| $70^0$ | 86 | 72 ± 5 | 1.14 | 84 | 70 ± 4 | 1.12 |
| $90^0$ | 37 | 20 ± 8 | 0.81 | 30 | 19 ± 6 | 0.72 |